\documentstyle[aps,twocolumn]{revtex}
\begin{document}
\title{Bose-Einstein Condensates with Large Number of Vortices}
\author{Tin-Lun Ho}
\address{Department of Physics,  The Ohio State University, Columbus, Ohio
43210}

\maketitle

\begin{abstract}
We show that as the number of vortices in a {\em three} dimensional 
Bose-Einstein Condensate increases, the system reaches a ``quantum Hall" 
regime where the density profile is a Gaussian in the $xy$-plane and an
inverted parabolic profile along $z$. The angular momentum of the system
increases as the vortex lattice shrinks. However, Coriolis force prevents the 
unit cell of the vortex lattice from shrinking beyond a minimum size. 
Although the recent MIT experiment is not exactly in the quantum Hall 
regime, it is close enough for the present results to be used as a guide. 
The quantum Hall regime can be easily reached by moderate changes of 
the current experimental parameters. 
\end{abstract}

In a recent experiment\cite{MIT}, the MIT group led by Wolfgang Ketterle has
generated as many as 160 vortices in a Bose-Einstein condensate of Na atoms. 
While the vortices form a hexagonal lattice as expected for equilibrium 
systems, many observed properties appear to differ
significantly from the predicted ones\cite{MIT}. It is thought
that the system may not be in equilibrium.  

Equilibrium or not, it is clear from the recent 
Paris\cite{Paris} and MIT experiments\cite{MIT} that one can deposit 
a large amount of angular momentum to the condensate by driving it with 
an asymmetric potential at a rotation frequency $\Omega^{}_{d}$ close to 
the quadrupolar resonance. In view of the recent rapid 
developments\cite{MIT}\cite{Paris}, it is conceivable that higher, 
or even much higher  angular momentum states can be achieved 
in the near future.  The high 
angular momentum states are interesting because they resemble  some of the
novel solid state systems such as type-II superconductor and  quantum Hall
liquid. However, the unique features of Bose-Einstein condensates also
introduce important differences. 

The purpose of this paper is discuss the properties of Bose-Einstein
condensates with large number of  vortices. We show that the physics in this
regime is closely related to two dimensional quantum Hall physics,
even though the system is three dimensional. The specific 
wavefunction in this regime allows one to calculate many properties of the
system, including the effective rotational frequency 
$\Omega = \partial E/\partial L$. It is clear that $\Omega$ will not be 
the same as the driving frequency $\Omega^{}_{d}$, since it
depends only on the total angular momentum $L$, which can be varied by
varying the duration of the drive, even though $\Omega^{}_{d}$ is held fixed. 
The fact that the vortex lattice has the equilibrium
form strongly suggests that the system is in quasi-equilibrium, characterized
by an $L$ and $\Omega$ which change slowly because of the residual asymmetry in
the confining potential.  It is conceivable that some of the discrepancy with
theory mentioned in ref.\cite{MIT} might disappear once the effective
frequency $\Omega$ is identified. 

We shall see that in the quantum Hall regime, Coriolis force 
prevents the unit cell $(v)$ of the vortices lattice
from shrinking below $v^{\ast}= \pi a^{2}_{\perp}$ where $a^{}_{\perp}$ is the
oscillator length in the $xy$-plane. For a large condensate, the density
profile along $z$ is an inverted parabola  similar to the stationary case.
However, it is a Gaussian in the $xy$-plane, with a width
$\sigma$ that scales as $R^{-3/2}_{z}$,
where $R^{}_{z}$ is the maximum extent of the condensate along $z$. The angular
momentum of the system is proportional to  $\pi \sigma^2$, which grows as the
size of unit cell $v$ shrinks. 

{\bf The quantum Hall regime and the minimum vortex separation}:
The condensate wavefunction $\Psi$ is determined by  the 
Gross-Pitaevskii functional
\begin{equation} 
{\cal K} = \int \Psi^{\ast}[ h^{}_{z} + h^{}_{\perp} - \Omega L^{}_{z} - \mu 
] \Psi  + \frac{1}{2}g\int |\Psi|^{4},  
\label{GP} \end{equation}
where $h^{}_{z}=
-\frac{\hbar^{2}}{2M}\nabla^{2}_{z} + \frac{1}{2}M\omega^{2}_{z} z^{2}$ and 
$h^{}_{\perp}= -\frac{\hbar^{2}}{2M}\nabla^{2}_{\perp} + 
\frac{1}{2}M\omega^{2}_{\perp} r^{2}$ are the single particle Hamiltonians 
along $z$ and in the $xy$-plane, $\omega^{}_{z}$ and $\omega^{}_{\perp}$ are
the frequencies of the harmonic potentials in $z$ and in the $xy$-plane;
${\bf r}\equiv  (x,y)$, 
$L^{}_{z} = -i \hbar \hat{\bf z} \cdot {\bf r}\times \nabla $ 
is the angular momentum; $g= 4\pi
\hbar^{2}a^{}_{sc}/M$, $a^{}_{sc}$ is the s-wave scattering length,
and ($\Omega$, $N$) are Lagrange multipliers specifying the angular
momentum and particle number of the system. 
It is useful to rewrite $h^{}_{\perp} - \Omega L^{}_{z}$
$={\cal H}_{L} +  \hbar (\omega^{}_{\perp} - \Omega) L^{}_{z}$, where
\begin{equation}
{\cal H}_{L} = \left( -i\hbar \nabla^{}_{\perp} - M\omega^{}_{\perp}
\hat{\bf z} \times  {\bf r} \right)^2/(2M) 
\label{H1} \end{equation}
Eq.(\ref{H1}) is precisely the Hamiltonian  $\frac{1}{2M} (-i\hbar
\nabla - e{\bf A}/c  )^2$ of a charge $-e$ particle moving in $xy$-plane 
subjected to a magnetic field $B\hat{\bf z}$ with a vector potential ${\bf A}=
\frac{1}{2}B\hat{\bf z}\times {\bf r}$, $eB/Mc = 2\Omega$.  
The eigenfunctions of ${\cal H}_{L}^{}$ are
\begin{equation}
h_{n,m}({\bf r}) = \frac{e^{|u|^2/(2a^{2}_{\perp})} 
\partial_{+}^{m}\partial_{-}^{n}
 e^{-|u|^2/a^{2}_{\perp}} }{\sqrt{ \pi a^{2}_{\perp} n! m!}}
\label{uf} \end{equation}
with eigenvalues $\epsilon^{}_{n,m} =\hbar\omega^{}_{\perp}(2n + 1)$ 
where $n$ and $m$ are integers $0, 1, 2, ..$, 
$u\equiv (x+iy)/a^{}_{\perp}$, $a^{}_{\perp}=\sqrt{\hbar/M\omega^{}_{\perp}}$,
$\partial_{\pm} \equiv (a^{}_{\perp}/2)(\partial_{x} \pm i \partial_{y})$.
The integer $n$ is referred to as the Landau level index, and $m$ labels the
degenerate states within a Landau level $n$. 
Since $u^{}_{n,m}$ are also eigenstates of 
$L^{}_{z}= \hbar (u\partial^{}_{u} - u^{\ast} \partial^{}_{u^{\ast}})$ 
with eigenvalue $\hbar(m-n)$.  The eigenvalue of 
$h_{\perp}^{} - \Omega L^{}_{z}$
is $\hbar [ (\omega^{}_{\perp} + \Omega) n +  (\omega^{}_{\perp} -
\Omega) m + \omega^{}_{\perp}]$. The degeneracy of the Landau level is lifted
by the deviation $\omega^{}_{\perp} -  \Omega$.\cite{typeII}  

Interaction effects will mix different $(n,m)$ states. For 
sufficiently small $g$ and $\omega^{}_{\perp} - \Omega$, the system primarily
resides in the lowest Landau level (LLL), ($n=0$), since other levels 
are separated from it by an energy gap $\geq \hbar \omega^{}_{\perp}$. 
At first sight, this might appear to be difficult to achieve for it requires 
$gn<< \hbar \omega^{}_{\perp}$, which implies 
either $g$ is exceedingly small or density the $n$ is very low\cite{Gunn}. 
While this is
true for low angular momentum states, it is not true in the high angular
momentum limit. As we shall see, as the number of vortices increases,
(corresponding to
$\Omega$ very close to $\omega^{}_{\perp}$), the density is thinned out
sufficiently in the $xy$-plane so that the LLL regime can be
attained. We shall therefore proceed by  assuming that the system in the 
LLL, and return later to show that mixing of higher 
Landau level is indeed small in the  fast rotating limit for large
condensates with  typical interactions.

Before proceeding, we note that since the dynamics along $z$ 
is identical to the  non-rotating case, we can apply the usual
Thomas-Fermi approximation along  $z$ to ignore the kinetic energy
$|\nabla^{}_{z}\Psi|^2$ term and justify  it at the end. Our task is then to
minimize $K = \int {\rm d}z {\cal K}(z)$,
\begin{equation}
{\cal K}(z) = \int {\rm d}{\bf r} \Psi^{\ast} \left[ 
h^{}_{\perp} - \Omega L^{}_{z} - \mu(z) \right] \Psi
  + \frac{1}{2}g\int |\Psi|^4 
\label{Kz} \end{equation}
with the constraint $N= \int |\Psi|^2$, where $\mu(z) = \mu - \frac{1}{2} M
\omega^{2}_{z} z^{2}$. 
Since $u_{0,m}({\bf r}) \propto u^{m}e^{-r^2/2a^{}_{\perp}}$, 
a general wavefunction function in the lowest Landau level is 
$\phi({\bf r}) = f(u) e^{-r^2/2a^{}_{\perp}}$, where $f(u)$ is an
analytic function of $u$.  Because of the Guassian cutoff, 
it is sufficient to take $f$ as a polynomial of power $Q$, which can be 
written in a factorized form $\prod^{Q}_{\alpha =1}(u - b^{}_{\alpha})$
according to the fundamental theorem of algebra, where $b^{}_{\alpha}$
are the zeros. 
 Since $f$ undergoes a $2\pi$ phase change as $u$
encircles $b^{}_{\alpha}$, the factor $(u - b^{}_{\alpha})$ describes a vortex
at ${\bf b}^{}_{\alpha}$, $b^{}_{\alpha}= (\hat{\bf x} + i \hat{\bf y})\cdot
{\bf b}^{}_{\alpha}$. If $Q$ extends to infinite, $f$ will be
an infinite product. The most general form of 
$\Psi$ within the lowest Landau level is then 
$\Psi(x,y, z) = 
C(z) \prod_{\alpha}(u- b^{}_{\alpha}(z))
e^{-r^{2}/(2a^{2}_{\perp})}$, or 
\begin{equation}
\Psi(x,y,z) = f(z)\phi(x,y; z), \,\,\,\,\,\,\,\, 
\int {\rm d}{\bf r} |\phi|^2 =1 
\label{PsiLLL} \end{equation}
where $\phi(x,y;z) = \overline{\phi}(x,y; z)/D(z)$, 
\begin{equation}
\overline{\phi}(x,y; z) =   e^{-r^{2}/(2a^{2}_{\perp})}
\prod_{\alpha =1}(u - b^{}_{\alpha}(z)), \,\,\,\,\,\,\,\,
\label{phibar} \end{equation}
and $D^{2} = \int {\rm d}{\bf r} |\overline{\phi}|^2$ is the normalization
constant for $\phi$.  The number constraint now becomes $\int {\rm d}z |f|^2
= N$.  Noting that within LLL, 
\begin{equation}
\int \Psi^{\ast}L^{}_{z} \Psi = \hbar \int
[(r/a^{}_{\perp})^2 -1] |\Psi|^2, 
\label{L} \end{equation}
and eq.(\ref{Kz})  becomes 
\begin{equation}
{\cal K}(z) =\left(- \tilde{\mu}(z) + 
\hbar (\omega^{}_{\perp} - \Omega)  \langle r^2\rangle_{\phi}^{}\right) f^2 +  
  \frac{g}{2} I_{\phi}^{} f^4 
\label{KKz} \end{equation}
where $\tilde{\mu}(z) = \mu - \hbar\Omega -
\frac{1}{2}M\omega^{2}_{z}z^2$, $\langle r^2\rangle_{\phi}^{}=
\int r^{2}  |\phi|^2 {\rm d}{\bf r}$ and 
$I_{\phi}^{} = \int {\rm d}{\bf r} |\phi|^4$. The minimization  is 
performed by varying $\{ b^{}_{\alpha} \}$. 
To evaluate $\langle r^2\rangle$ and $I_{\phi}^{}$, we note that 
\begin{equation}
|\overline{\phi}|^2 = e^{-{\cal H}}, \,\,\,\,\,\,
{\cal H} = r^{2}/a^{2}  - 2\sum_{\alpha}{\rm ln}
|{\bf r} - {\bf b}^{}_{\alpha}|. 
\label{SS} \end{equation}
The quantity ${\cal H}$ in eq.(\ref{SS}) is precisely the energy of a
charge $q=-1$ particle in two dimension interacting with a uniform positive
charge background $\rho^{}_{b} = \pi/a^{2}$, and a 
set of $q = -1$ negative charges located at $\{ {\bf b}^{}_{\alpha} \}$. 
From eq.(\ref{SS}), we have 
\begin{equation}
\nabla^{2}_{\perp} {\cal H} = 4 \pi \left(  (\pi a^2)^{-1} -
  \sum_{\alpha}\delta({\bf r} - {\bf b}_{\alpha})  \right) ,
\label{Gauss} \end{equation}
which is the Gauss's law for the 2D charged system. 
If $\{ {\bf b}^{}_{\alpha} \}$ forms an infinite regular lattice, 
${\bf b}^{}_{n^{}_{1}, n^{}_{2}}
= n^{}_{1} {\bf c}_{1}^{} + n^{}_{2} {\bf c}_{2}^{}$, ($n^{}_{1}, n^{}_{2}$
integers), then 
$\sum_{\alpha}\delta({\bf r} - {\bf b}^{}_{\alpha})  = 
v^{-1}\sum_{{\bf K}} e^{i{\bf K}\cdot {\bf r}}$, 
where $v= |{\bf c}^{}_{1}\times {\bf c}^{}_{2}|$ 
is the size of the unit cell, and ${\bf K}= \ell^{}_{1}{\bf
K}^{}_{1} + \ell^{}_{2}{\bf K}^{}_{2}$ ($\ell_{1}^{}, \ell^{}_{2}$ integers) 
are the reciprocal lattice vectors; (${\bf K}_{1}= (2\pi/v){\bf
c}_{2}\times\hat{\bf z}$, ${\bf K}_{2}= (2\pi/v)\hat{\bf z}\times {\bf
c}_{1}$). Eq.(\ref{Gauss}) and (\ref{SS}) then become
\begin{equation}
\nabla^{2}_{\perp} {\cal H} =  
\frac{4}{\sigma^2} - \frac{4\pi}{v}  \sum^{}_{{\bf K}\neq 0} 
{\rm cos}{\bf K}\cdot {\bf r}  , \,\,\,\,\,
\frac{1}{\sigma^{2}} = \frac{1}{a^{2}_{\perp}} - \frac{\pi}{v} , 
\label{sss} \end{equation}
\begin{equation}
|\overline{\phi}|^2 = e^{- (r/\sigma)^2}  \prod^{}_{{\bf K}\neq 0} 
e^{ - \zeta^{}_{\bf K}{\rm cos}{\bf K}\cdot {\bf r}}, 
\,\,\,\,\,\,\,
\zeta^{}_{K} = \frac{4\pi}{v {\bf K}^2} . 
\label{newphi} \end{equation}
The Gaussian in eq.(\ref{newphi}) corresponds to replacing the vortex
density  $\sum^{}_{\alpha}\delta({\bf r} - {\bf b}^{}_{\alpha})$ in 
eq.(\ref{Gauss}) by its average  $1/v$. In the 2D electrostatics analog, it
corresponds to reducing the uniform positive background by the average
density of the discrete negative charges. Since the renormalized background is
 less confining for the negative unit charge at $u$, we have $\sigma>
a^{}_{\perp}$. Eq.(\ref{sss}) also requires  $v> v^{\ast} = \pi a^{2}_{\perp}$,
for otherwise $\phi$ will not be normalized.  For hexagonal lattice, this
means that the distance $c$ between neighboring vortices can not be  shorter
than
$(2\pi/\sqrt{3})^{1/2}a^{}_{\perp}$. 

The angular dependence of $|\overline{\phi}|^2$ is contained in the product in 
eq.(\ref{newphi}). Since $v {\bf K}^2$ does not depend on the length of the
basis vector ${\bf c}^{}_{\alpha}$ but only on the lattice type (i.e. square or
hexagonal), the product in eq.(\ref{newphi}) determines mainly  the
lattice structure, and is less important in determining the overall density
profile of the system.   We shall therefore proceed by replacing the
vortex density in eq.(\ref{Gauss}) by its average, (referred to as ``averaged 
vortex approximation", and come back later to consider the effect of the
product in eq.(\ref{newphi}). We shall see that the results of the
averaged vortex approximation are intact except for changes of numerical
factors (of order 1).

{\bf Averaged  vortex approximation}: Keeping only the Guassian in
eq.(\ref{newphi}),
the {\em normalized} function $\phi$ in eq.(\ref{PsiLLL}) becomes  
$\phi({\bf r}) = \frac{1}{\pi \sigma^2}e^{-r^2/\sigma^2}$. Eq.(\ref{KKz}) then
becomes
\begin{equation}
{\cal K}(z) = - \left[ \tilde{\mu}(z)  -
\frac{ \hbar ( \omega^{}_{\perp} - \Omega) }{ (a^{}_{\perp}/\sigma)^2 }
\right] f^2 +  
 \frac{\hbar \omega^{}_{\perp} a^{}_{sc} f^4 }{ (\sigma/a^{}_{\perp})^2 } .
\label{KKKz} \end{equation}
The optimal $\sigma$ and $f$ are 
\begin{equation}
\left[ \frac{\sigma(z)}{a^{}_{\perp}} \right]^2 =
\frac{\tilde{\mu}(z) }
{3\hbar (\omega^{}_{\perp} -  \Omega) } , \,\,\,\,
a^{}_{sc} f^{2}(z) = \frac{
[\tilde{\mu}(z)]^2}{ 9 \hbar^2 \omega^{}_{\perp} (\omega^{}_{\perp} - 
\Omega) } .
\label{ascf2} \end{equation}
Since $\sigma\leq a^{}_{\perp}$, the maximum extent of the cloud along $z$
satisfying this condition (denoted as $R^{}_{z}$) is given by 
$3\hbar (\omega^{}_{\perp} - \Omega) = \tilde{\mu}(R^{}_{z})$.  Note that  
within the average vortex approximation, $f(z)$ must vanish for 
$|z|>R^{}_{z}$ in order to keep $\Psi$ normalized. 
We then have 
\begin{equation}
\tilde{\mu}(z) = \frac{1}{2}M \omega^{2}_{z}(R^{2}_{z} - z^{2}) + 3
\hbar(\omega^{}_{\perp} - \Omega) . 
\label{mutilde} \end{equation}
The number condition $N= \int f^{2} {\rm d}z$ then gives
\begin{equation}
\frac{R^{}_{z}}{a^{}_{z}}  =  \left[ \frac{135 N}{4}
\left(\frac{a^{}_{sc}}{a^{}_{z}}\right) \left(
\frac{\omega^{}_{\perp}}{\omega^{}_{z}} \right)^2 \left( 1-
\frac{\Omega}{\omega^{}_{\perp}}\right) \right]^{1/5}
(1 +  ..) 
\label{rovera} \end{equation}
where $(1+ ...) = ( 1 - 3x + O(x^2)
 + .. )$,  $x = [(\frac{\omega^{}_{\perp}}{\omega^{}_{z}})^{1/5}
(1-\frac{\Omega}{\omega^{}_{\perp}})^{3/5}]
/[\frac{135 N}{4} \frac{a^{}_{sc}}{a^{}_{z}} ]^{2/5}$. As we shall see, $x<<1$
for typical experimental parameters and can be ignored. 
From eq.(\ref{sss}) and (\ref{ascf2}), the size of the unit cell is given by 
\begin{equation}
v(z)  = \pi a^{2}_{\perp} \left(1 
+ \frac{6 (\omega^{}_{\perp} - \Omega)}{\omega^{}_{z} }
\frac{ a^{2}_{z}}{R^{2}_{z} - z^2} \right) . 
\label{vz} \end{equation}
In the fast rotating limit, 
$\Omega \rightarrow \omega^{}_{\perp}$, 
eq.(\ref{vz}) is essentially the usual relation $n_{v}= 2\Omega/(h/M)$ 
for estimating the vortex density but in a more detailed form. (The connection
is easily seen since $n_{v} = 1/v$, and $\pi a^{2}_{\perp} =
\pi M\omega^{}_{\perp}/\hbar \approx 2M\Omega/h$. 
 Note also that even though $\sigma^2$ shrinks quadratically as
$z$  increases (eq.(\ref{ascf2})), $v(z)$ (and hence the spacing between
vortices)  changes very little except near the tips of the condensate 
at $z \approx \pm R^{}_{z}$. 

A useful quantity is $(\pi\sigma^{2}/v)^{}_{o}\equiv [\pi\sigma^{2}/v]_{z=0}$, 
the number of vortices inside an area $\pi \sigma^{2}$ at the center cross 
section (i.e. $z=0$). From eq.(\ref{ascf2}) and (\ref{vz}), we have 
\begin{equation}
\left(\frac{\pi \sigma^2}{v}\right)_{o} = \left( 
\frac{ \pi a^{2}_{\perp}}{v - \pi a^{2}_{\perp}} \right)_{o}
= \frac{\omega^{}_{z}}{6 (\omega^{}_{\perp} - \Omega)} 
\left(  \frac{R^{2}_{z}}{a^{2}_{z}}\right) 
\label{pi} \end{equation}
Combining eq.(\ref{pi}) and eq.(\ref{rovera}), we can express the equilibrium
frequency $\Omega$ and $R^{}_{z}$  in terms of the number of vortices in the
center cross section, 
\begin{equation}
\left( 1- \frac{\Omega}{\omega^{}_{\perp}}\right) 
= \left(\frac{v}{6\pi \sigma^2} \right)^{5/3}_{o}
\left( \frac{\omega^{}_{z}}{ \omega^{}_{\perp} } \right)^{1/3} 
\left[ \frac{135N}{4}\left(\frac{a^{}_{sc}}{a^{}_{z}} \right) \right]^{2/3}, 
\label{subcri} \end{equation}
\begin{equation}
\frac{R^{}_{z}}{a^{}_{z}}  =  \left[ \frac{135 N}{4}
\left(\frac{a^{}_{sc}}{a^{}_{z}}\right) \left(
\frac{\omega^{}_{\perp}}{\omega^{}_{z}} \right) \left(\frac{v}{6\pi \sigma^2}
\right)_{o} \right]^{1/3}+ .. 
\label{newrovera} \end{equation} 
From eq.(\ref{ascf2}) and (\ref{rovera}), we find that the 
\begin{equation}
\sigma^2_{o}/ a^{2}_{\perp} = 
(45N/8) (a^{}_{sc}/a^{}_{z})(\omega^{}_{\perp}/\omega^{}_{z})
 (a^{}_{z}/R^{}_{z})^{3}
\end{equation}
The quantities $\Omega$ and $R^{}_{z}$ for systems with different 
particle number and vortex numbers are shown in Table I. 
Finally, we note that the total angular momentum eq.(\ref{L}) 
of the system is $\langle L^{}_{z}\rangle =  \hbar\frac{6N}{7}\left(
\frac{\pi \sigma^2}{v} \right)$. The total energy is $E = 
{\cal K} + \Omega \langle L_{z}^{}\rangle + \mu N$. To the leading term in $N$,
we have 
$E = \frac{\omega^{}_{\perp}}{243}$
$\left(\frac{\omega^{}_{z}}{\omega^{}_{\perp}}\right)^3$
$\left(1- \frac{\Omega}{\omega^{}_{\perp}}\right)^{-2}$
$\left(\frac{a^{}_{z}}{a^{}_{sc}}\right)(R^{}_{z}/a^{}_{z})^{7}$, hence 
$E \propto N^{7/5}$. 

{\em Mixing of higher Landau levels and the validity of Thomas-Fermi
approximation (TFA)}: The validity of TFA requires requires $(\hbar^2/2M)
|\nabla_{z}^{}
\Psi|^2 << \mu(z) |\Psi|^2$, which is $\hbar^{2}/(2MR^{2}_{z})<< \mu $
or $(R^{2}/a_{\perp}^{})^4 >>1$. This is  easily satisfied for the
range of parameters listed in Table I.  
Mixing of higher Landau level will be
unimportant if the interaction energy density 
$\hbar\omega^{}_{\perp} (a^{}_{\perp}/\sigma)^2 a^{}_{sc} f^4$ is much less
than that of higher Landau level $\hbar\omega^{}_{\perp}f^2$, or 
$\Gamma \equiv   (a^{}_{\perp}/\sigma)^2 a^{}_{sc} f^2 <<1$. From
eq.(\ref{ascf2}), we  have 
$\Gamma = \tilde{\mu}(z=0)/(3\hbar\omega^{}_{\perp})$, or 
$\Gamma =\frac{\omega^{}_{z}}{6\omega^{}_{\perp}} \left( 
R^{}_{z}/a^{}_{z}\right)^2$. The values of $\Gamma$ for a variety of external
parameters are shown in Table I. Although the recent MIT experiment\cite{MIT} 
is not yet in the LLL regime, it is not too far away. It is clear from Table I
that the LLL regime can be reached within the capability of current technology. 
Note also that as $\Omega \rightarrow
\omega^{}_{\perp}$ and the system is driven more toward the LLL regime. However,
$R_{z}^{}/a^{}_{z}$ also decreases. As 
$R_{z}^{}$ becomes very close to $a^{}_{z}$, Thomas-Fermi 
approximation will not be
valid and a more accurate treatment along $z$ is needed. The present treatment
is valid for the frequencies $\Omega$ such that $R^{}_{z}/a^{}_{z}>1$, which is
satisfied for the cases listed.

{\bf Beyond averaged vortex approximation}: A full calculation of the
quantities $\langle r^2 \rangle_{\phi}^{}$ and $I_{\phi}^{}$ in 
eq.(\ref{Kz}) requires considering the ``structural" product in
eq.(\ref{newphi}). We shall show that in the limit $\Omega \rightarrow
\omega^{}_{\perp}$, the major effect of the structural
product is to stabilize the hexagonal lattice over the square lattice. 
It does not alter the result 
$\langle r^2 \rangle_{\phi}^{}$ of the 
averaged vortex approximation, but changes $I_{\phi}$ from $1/(2\pi
\sigma^2)$ to $\alpha/(2\pi \sigma^2)$ with $\alpha \sim 1$. Thus, all 
results of the simple average vortex approximation remain. 

To prove the above statements, we recall that 
$e^{-\alpha {\rm cos}\theta} = \sum_{n=0, 1, 2, ..} (-1)^{n}
I_{n}^{}(\alpha) e^{in \theta}$, where $I^{}_{n}(\alpha)$ are the
modified Bessel functions. Eq.(\ref{newphi}) then becomes 
$|\overline{\phi}|^2 =  e^{-r^{2}/\sigma^2} \prod^{}_{{\bf K}\neq 0}
\left(
\sum^{}_{n_{\bf K}} (-1)^{n_{\bf K}} I_{n_{\bf K}}  (\zeta^{}_{\bf K})
e^{in_{\bf K} {\bf K}\cdot {\bf r}} \right)$, or 
\begin{equation}
|\overline{\phi}|^2 =  e^{-r^{2}/\sigma^2}\sum_{[n_{\bf K}]} ' 
 \Lambda [n_{\bf K}, \zeta^{}_{\bf K}]
e^{i{\bf P}[n_{\bf K}]\cdot {\bf r}} , 
\label{sum'}\end{equation}
\begin{equation}
\Lambda [n_{\bf K}, \zeta^{}_{\bf K}] = \prod_{{\bf K} \neq 0} (-1)^{n_{\bf K}}
I_{n_{\bf K}}(\zeta^{}_{\bf K}), \,\,\,\,\,\,
{\bf P}[n_{\bf K}] = \sum_{\bf K} {\bf K} n_{\bf K}, 
\end{equation}   
where  $\sum'$ in eq.(\ref{sum'})  means the ${\bf K}=0$ term is excluded, 
and $[n_{\bf K}, \zeta^{}_{\bf K}]$ denotes the entire sets $\{ n_{\bf K}^{}
\}$ and $\{ \zeta_{\bf K}^{} \}$. 
The {\em normalized} function $|\phi|^2$ is then
\begin{equation}
|\phi|^{2} =  \frac{e^{-r^{2}/\sigma^2} }{ \pi \sigma^2} 
\frac{ \prod_{{\bf K} \neq 0}  e^{-\zeta_{\bf K} {\rm cos}{\bf K}\cdot {\bf
r}}} 
{Z_{\sigma^2} [ \zeta_{\bf K} ]} , 
\label{norphi} \end{equation}
\begin{equation}
Z_{\sigma^2} [ \zeta_{\bf K} ] =  \sum_{[n^{}_{\bf K}]}
\Lambda [n_{\bf K}, \zeta^{}_{\bf K}]
e^{ -\sigma^{2}P[n^{}_{\bf K}]^{2}/4}. 
\label{Z} \end{equation}
We then have $\langle r^{2}\rangle_{\phi} = \sigma^2 \gamma$, 
$\int |\phi|^4 = \alpha /(2\pi \sigma^2)$, where 
$\gamma=$
$Z_{\sigma^2} [ \zeta_{\bf K} ]^{-1}$
$\sum_{[n^{}_{\bf K}]}\Lambda [n_{\bf K}, \zeta^{}_{\bf K}] $
$e^{ -(\sigma P[n^{}_{\bf K}]/2)^{2} } $
$\left(1- \frac{\sigma^2 P[n^{}_{\bf K}]^{2}}{4}\right)$, and 
$\alpha = Z_{\sigma^2/2} [ 2\zeta_{\bf K} ]$
$Z_{\sigma^2} [ \zeta_{\bf K} ]^{-2}$. 
As $\Omega\rightarrow \omega^{}_{\perp}$, $\sigma^{2}P[n_{\bf K}]^2>>1$ 
for all ${\bf P}[n_{\bf K}]\neq 0$. Only terms in eq.(\ref{Z}) 
with $\sum_{\bf K} {\bf K}n_{\bf K} =0$ are important. 
This implies $\gamma=1$, $\langle r^{2} \rangle_{\phi} = \sigma^2$, and 
\begin{equation}
\alpha =  \tilde{Z}[2\zeta_{\bf K}] / ( \tilde{Z}
[ \zeta_{\bf K} ]) ^2 , 
\label{J4} \end{equation}
\begin{equation}
\tilde{Z}[\zeta_{\bf K}]
= \sum_{n_{\bf K}}'' \prod_{{\bf K}\neq 0} (-1)^{n_{\bf K}} \tilde{I}_{n_{\bf
K}}(\zeta_{\bf K}), 
\label{Ztilde} \end{equation}
where $\sum''$ means summing over $\{ n_{\bf K}^{} \}$ with 
such that  $\sum_{\bf K} {\bf K}n_{\bf K}=0$. 

To evaluate $\alpha$, we 
rewrite $\tilde{Z}$ as 
$\tilde{Z}[\zeta_{\bf K}] = W[\zeta_{\bf K}]
\sum_{[n^{}_{\bf K}]}''  \prod_{{\bf K}\neq 0} (-1)^{n_{\bf K}} 
\tilde{I}_{n_{\bf K}}(\zeta_{\bf K}) $, 
where $W[\zeta_{\bf K}] = \prod_{{\bf K}\neq 0} I_{0}(\zeta_{\bf K})$, and 
$\tilde{I}_{n}(\zeta) = I_{n}(\zeta)/ I_{0}(\zeta)$. 
Let us define $\zeta^{}_{o}= \zeta^{}_{{\bf K}_{o}}$, where ${\bf
K}_{o}$ is the shortest reciprocal lattice vector. Since $\zeta^{}_{o} =
\sqrt{3}/(2\pi)$ and $1/\pi$
for hexagonal and square lattice respectively, and since
$I_{n}(\zeta) = \sum_{k=0}^{\infty} \frac{(\zeta/2)^{n+2k}}{k! (n+k)!}$,
the sums in eq.(\ref{J4}) are fast convergent. 
Regarding $\tilde{Z}$ formally as a power series of $\zeta^{}_{o}$, 
we have for the hexagonal lattice, 
$W[\zeta_{\bf K}] = 1 + (3\zeta^{2}_{o}/2) \beta_{1}^{}+ ..$, and 
\begin{equation}
\tilde{Z}[ \zeta_{\bf K} ] = W[\zeta_{\bf K}]
\left( 1 + \frac{3 \zeta^{2}_{o}}{4} \beta_{1} 
- \frac{ \zeta^{3}_{o}}{4} \beta_{2} + ... \right)
\label{hexa} \end{equation}
where $\beta_{n} = \sum_{\ell_{1}=1}^{\infty}\sum_{\ell_{2}=0}^{\infty}
(\ell^{2}_{1} + \ell^{2}_{2} + \sqrt{3} \ell_{1}\ell_{2})^{-2n}$. The
$\zeta^{2}_{o}$ term in $W$ comes from expanding 
$I^{}_{o}(\zeta_{\bf K})$ in $\zeta^{}_{o}$ to second order in $\zeta^{}_{o}$.
 The $\zeta^{2}_{o}$ term in eq.(\ref{hexa})
comes from pairs of $\tilde{I}^{}_{1}$ terms in $\tilde{Z}$. The factor 
$\beta_{1}^{}$ is the result of summing over all such pairs. 
The $\zeta^{3}_{o}$ term in eq.(\ref{hexa}) comes from triplets of vectors of
the same length in eq.(\ref{Ztilde}). These 
triplets give rise to a {\em negative}
sign which lowers the energy of the system. 
The factor $\beta_{2}^{}$ is the result of
summing all these isosceles triplets. Using eq.(\ref{hexa}) and eq.(\ref{J4}), 
we have $\alpha =  1 + \frac{9}{2} \zeta^{2}_{o}\beta^{}_{1}
- \frac{1}{4} \zeta^{3}_{o} \beta^{}_{2}+ ...$ to orders less than
$\zeta^{4}_{o}$.  The result is $\alpha = 1.38$, ($\beta_{1} = 1.218$,
$\beta_{2} = 1.001$,
$\zeta^{}_{o} = 0.276$).

Repeating the same calculation for the the square
lattice. We have, to the lowest order in $\zeta^{}_{o}$, 
$W[\zeta_{\bf k}] = 1 + \zeta^{2}_{o} \gamma_{1}+ ..$, $\tilde{Z}[\zeta_{\bf
K}] = W[\zeta_{\bf k}] ( 1 + \frac{1}{2} \zeta^{2}_{o} \gamma_{1} + ...)$,
where $\gamma_{1}$$ = $$\sum_{\ell_{1} =1}^{\infty}$
$\sum_{\ell_{2} = 0}^{\infty}$
$(\ell^{2}_{1} + \ell^{2}_{2})^{-2} = 1.506$, 
which gives $\alpha = 1 + 3 \zeta^{2}_{o}\gamma_{1}+..$ to orders less than
$\zeta^{4}_{o}$. The result is $\alpha= 1.45$,  since $\zeta^{}_{o} = 0.318$.  
The square lattice has energy higher than the hexagonal one because it 
has no negative triplets terms.

{\em Final Remarks:} We have shown that as the number of vortices increases, 
a Bose-Einstein condensate will become quantum Hall like.  
A natural question is what happens if the angular 
momentum of the system keeps increasing. 
We have seen that $R^{}_{z}$ decreases as angular
momentum increases. Eventually, it will reduce to the point where only one 
harmonic state in the z-direction will be occupied. In such ``ultra-fast"
rotating limit,
fluctuations effects in the $xy$ plane will be important and one has to 
go beyond mean field treatment. The discussion of the ultra-fast limit
will be presented elsewhere. 

I thank Wolfgang Ketterle for sending me his preprint before its publication, 
and Mehmet Oktel for discussions. 
This work is supported by the NASA Grant
NAG8-1441, and the NSF Grants DMR-9807284, DMR-0071630. 

\vspace{0.1in}

\begin{tabular}{|l|l|l|l||l|l|l|l|} \hline
 &N$(10^6)$ & $\omega^{}_{\perp}/2\pi$ &
$\omega^{}_{z}/2\pi$ &
$\pi \sigma^{2}/v$ & $1-\Omega/\omega^{}_{\perp}$ & $R^{}_{z}/a^{}_{z}$ &
 $\Gamma$  \\ \hline
a & $50$ & 84 & 20 & 40 & 0.64 & 6.8 &
1.85 \\ \hline 
b & $50$ & 1000 & 10 & 40 & 0.179 &
10.9 & 0.198 \\ \hline 
c & $50$ & 1000 & 10 & 100 & 0.038 &
9.3 & 0.146 \\ \hline 
d & $5$ & 84 & 20 & 40 & 0.13 &
4.6 & 0.86 \\ \hline 
e & $5$ & 1000 & 10 & 40 & 0.039 &
7.4 & 0.092 \\ \hline 
f & $5$ & 1000 & 10 & 100 & 0.008 &
6.38 & 0.06 \\ \hline 
\end{tabular}

\noindent Caption: 
Row $(a)$ corresponds to the parameters of the recent MIT experiment\cite{MIT}.
We have counted the vortices as follows. 
The Gaussian density profile $|\phi|^2$ in eq.(\ref{newphi}) drops
by a factor of 100 from the center to a radius $r=2\sigma$. If there are 160
vortices within a radius of $2\sigma$, then $\pi\sigma^2/v = 40$. To remain 
in the lowest Landau level (LLL), one needs $\Gamma <<1$. The 
parameters of recent MIT experiment\cite{MIT} is not yet in the LLL regime. 
However, the latter can be reached by changing the parameters as indicated
in case (b) to (f). For all cases, the quantity $x$ defined after
eq.(\ref{rovera}) is $\sim 10^{-3}$.


\begin{thebibliography}{99}
\bibitem{typeII} At criticality, $\Omega=\omega^{}_{\perp}$, eq.(\ref{GP}) is
identical to the free energy of type-II
superconductor.  The two problems are different 
because the constraint $N = \int |\Psi|^2$ does not
exist in type-II superconductors, and the Maxwell equation 
in type-II superconductors does not exist in Bose gases.
\end{thebibliography}
\end{document}